\documentclass[fleqn,usenatbib]{mnras}
\usepackage{newtxtext,newtxmath}
\usepackage[T1]{fontenc}
\usepackage{ae,aecompl}

\usepackage{subfig,graphicx,caption}
\usepackage{amsmath}	    

\defcitealias{2020MNRAS.492.3294M}{\textsc{MB20}}
\defcitealias{2021MNRAS.503.2097M}{\textsc{paper-i}}
\defcitealias{2021arXiv210313617M}{\textsc{paper-ii}}
\defcitealias{2020arXiv201209994N}{N20}

\title[Exotic Image Formation in Cluster Lenses]
{Exotic Image Formation in Strong Gravitational Lensing by Clusters 
of Galaxies -- III: Statistics with HUDF}

\author[A. K. Meena and J. S. Bagla]{
Ashish Kumar Meena$^{1}$\thanks{E-mail: ashishmeena766@gmail.com},
Jasjeet Singh Bagla$^{1}$\thanks{E-mail: jasjeet@iisermohali.ac.in}
\\
\\
$^{1}$Indian Institute of Science Education and Research Mohali,
Knowledge City, Sector 81, SAS Nagar, Punjab 140306,
India}

\pubyear{2021}

\begin{document}
\label{firstpage}
\pagerange{\pageref{firstpage}--\pageref{lastpage}}
\maketitle

\begin{abstract}

We study the image formation near point singularities 
(swallowtail and umbilics) in the simulated strongly 
lensed images of Hubble Ultra Deep Field (HUDF) by the Hubble 
Frontier Fields (HFF) clusters. 
In this work, we only consider nearly half of the brightest
(a total of 5271) sources in the HUDF region.
For every HFF cluster, we constructed 11 realizations of 
strongly lensed HUDF with an arbitrary translation of the
cluster centre within the central region of HUDF and an 
arbitrary rotation. 
In each of these realizations, we visually identify the 
characteristic/exotic image formation corresponding to the 
different point singularities. 
We find that our current results are consistent with our
earlier results based on different approaches. 
We also study time delay in these exotic image formations 
and compare it with typical five-image geometries. 
We find that the typical time delay in exotic image formations
is an order of magnitude smaller than the typical time
delay in a generic five-image geometry.

\end{abstract}


\section{Introduction}
\label{sec: Introduction}

Galaxy clusters strong lenses are a unique probe 
to study the physics of the Universe 
\citep[e.g.,][]{1937PhRv...51..290Z, 1992ARA&A..30..311B, 
2011A&ARv..19...47K}.
The strongly (and weakly) lensed sources allows us to construct 
the mass models of these clusters, which further helps us 
understand the intracluster dynamics 
\citep[e.g.,][]{2019MNRAS.487.1905A, 2020arXiv201206611A}.
At the same time, these cluster lenses enables us to observe 
distant galaxies \citep[e.g.,][]{2013ApJ...762...32C}, 
multiply imaged supernova \citep{2015Sci...347.1123K}, 
and highly magnified stars \citep{2018NatAs...2..334K}.
Although at present the number of such systems is small but 
with the upcoming facilities like 
Euclid \citep{2009arXiv0912.0914L}, 
Nancy Grace Roman Space Telescope \citep[WFIRST;][]{2019arXiv190205569A}, 
James Webb Space Telescope \citep[JWST;][]{2006SSRv..123..485G}, 
Vera Rubin Observatory \citep[LSST;][]{2019ApJ...873..111I}, 
the number of such systems is expected to increase by more than an 
order of magnitude \citep[e.g.,][]{2010MNRAS.405.2579O, 2015ApJ...811...20C}.
With such a significant increase in strongly lensed systems, 
we also expect an increment in the observed number of image 
formations near point singularities.

Point singularities (e.g., swallowtail, hyperbolic and elliptic umbilics) 
are singularities of the lens mapping and only occur for specific 
source redshifts \citep[][hereafter \textsc{MB20}]{2020MNRAS.492.3294M}. 
These point singularities come with a characteristic image formation
and are very sensitive to the lens model parameters. 
Hence, the number of point singularities can be (significantly) 
different for different mass models of a cluster lens
\citep[][hereafter \textsc{paper-i}]{2021MNRAS.503.2097M}.
An earlier study \citep{2009MNRAS.399....2O} estimated that the 
observed number of image formations near these point singularities 
is expected to be very small in the full sky surveys. 
However, our recent study (\citetalias{2021MNRAS.503.2097M}) 
showed that we can expect to observe one image 
formation near swallowtail and hyperbolic umbilic in every five 
cluster lenses in the JWST era. 
These numbers denote lower limits even if we include the 
corresponding statistical uncertainties 
\citep[][hereafter \textsc{paper-ii}]{2021arXiv210313617M}.
These numbers of image formations near point singularities are
estimated by using the predicted source galaxy population (taken
from \citealt{2018MNRAS.474.2352C}) observed by 
JWST with an observation time of $10^4$ seconds.

Another way to estimate the number of image formations near
point singularities is to look for the characteristic image
formations in simulated strongly lensed regions of the sky
\citep[e.g.,][]{2016ApJ...828...54L, 2019MNRAS.482.2823P}. 
So far, such simulated sky regions have been used to compare 
different cluster mass reconstruction methods and study
systematic uncertainties \citep{2010A&A...514A..93M, 2017MNRAS.472.3177M}.
Such a method helps us to understand how well a reconstruction 
method can recover various lens properties and how one can make 
further improvements in the reconstruction method.
Understanding the lens mass distribution well, in turn, also 
helps in better understanding the intrinsic properties of the 
source \citep[e.g.,][]{2019A&A...623A..14J, 2020MNRAS.496.2648Y}.
Such a method can also allow us to validate the results of 
\citetalias{2021MNRAS.503.2097M} and \citetalias{2021arXiv210313617M},
by looking into the simulated strongly lensed sky images for 
characteristic image formations near point singularities.
As these characteristic image formations comprise multiple 
images in a very small region of the sky, these image formations 
can be very beneficial for time-delay cosmography 
\citep[e.g.,][]{2016A&ARv..24...11T}.

In current work, we study the possibility of 
observing characteristic image formation in simulated strongly 
lensed images of the sky. 
However, instead of simulating unlensed sky patch and then 
performing strong lensing, we lensed the Hubble Ultra Deep 
Field \citep[HUDF;][]{2006AJ....132.1729B} using the Hubble 
Frontier Fields \citep[HFF;][]{2017ApJ...837...97L} cluster lenses.
The HUDF is the deepest image of the sky, consisting of sources 
all the way up to (photometric) redshift ${\sim}8$, whereas the 
HFF clusters are some of the very well-studied galaxy cluster lenses.
The advantage of such an approach is that one does not have 
to worry about the k-corrections or the alignment of the source 
galaxies in the sky.
Apart from that, one also does not have to worry about the 
luminosity function of the source galaxies.
However, at the same time, one is limited by the patch size of 
the HUDF and cannot go beyond it, which gives rise to concerns 
related to cosmic variance \citep[e.g.,][]{2007ApJ...671.1212O, 
2010MNRAS.407.2131D, 2011ApJ...731..113M}.
And, as the HUDF is a small patch of the sky, the corresponding 
luminosity function cannot be generalized to the whole sky,
but, the combination of HUDF with other deep fields can 
lead to robust measurements of the galaxy luminosity function
\citep[e.g.,][]{2015ApJ...803...34B, 2015ApJ...810...71F, 
2016MNRAS.456.3194P}.
Hence, our results based on the HUDF lensing are probably affected 
by the cosmic variance, but due to the unprecedented depth of 
the HUDF, we proceed with it.
We make multiple lensing realizations for each cluster lens and 
visually identify the image formation near the point singularities. 
Such a visual inspection is possible due to the characteristic 
nature of these image formations (see 
\citetalias{2020MNRAS.492.3294M} for more details).
Here we use the best-fit \textsc{grale} mass models for the HFF
clusters.
As a result, we only focus on the swallowtail, and hyperbolic 
umbilic (purse) singularities as the number of elliptic umbilics 
(pyramid) in the \textsc{grale} mass models for the HFF clusters 
is negligible (see \citetalias{2021arXiv210313617M} for the HFF
cluster singularity maps).
Further, as highlighted in \citetalias{2021MNRAS.503.2097M}, 
use of \textsc{grale} mass maps gives us the most conservative
estimate of the number of point singularities. 

Apart from identifying these characteristic image formations, 
we also study the corresponding time delay distribution. 
In order to use these characteristic image formations for 
time-delay cosmography, one needs to check whether the corresponding 
time delays lie within a reasonable observation time frame.
In general, both swallowtail and hyperbolic umbilic (purse) 
come with a five-image geometry. Out of these five images, 
four lie very close to each other in the image plane in the 
form of an arc (for swallowtail) or a ring (for purse). 
As a result, the time delay between these images is expected to 
be small compared to a generic five-image geometry.
Hence, we also compare the time delay distribution in characteristic 
image formation with the time delay distribution in typical 
five-image geometry observed in the lensed HUDF templates.

The paper is organized as follows. 
Section \ref{sec:basics} briefly reviews the basics of 
gravitational lensing and singularities therein.
The HFF cluster lenses are enumerated in Section \ref{sec:hff_clusters}.
The relevant details of HUDF are revisited in Section \ref{sec:hudf}.
In Section \ref{sec:results}, we present our results.
Summary and conclusions are presented in Section \ref{sec:conclusions}.
We also discuss our future work in this section.

\section{Basic Gravitational Lensing}
\label{sec:basics}

\begin{figure*}
	\centering
	\includegraphics[height=5.7cm, width=17cm]{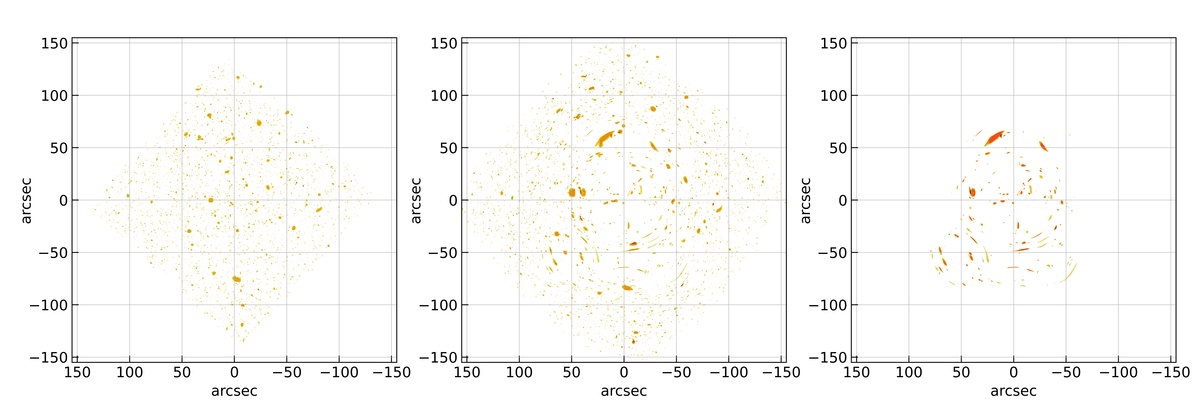}
	\caption{An example of magnification cut on the lensed HUDF: 
	The left panel represents the unlensed HUDF. The middle panel
	represent the lensed HUDF without any magnification cut. Here,
	the lens in A370 with center aligned with HUDF and no rotation,
	i.e., the zeroth realization. The right panel is same as middle
	panel but with lensed sources that satisfies $\mu{\geq}5$ 
	criteria.} 
	\label{fig:a370_real_0}
\end{figure*}

In this section, we briefly review the relevant basics of 
gravitational lensing and singularities therein.
For a more detailed discussion, we encourage reader to look into
\citet{1992grle.book.....S} and \citetalias{2020MNRAS.492.3294M}.

The lens equation which relates the unlensed source position and 
the corresponding lensed image position (in dimensionless form) 
is given as:
\begin{equation}
	\mathbf{y} = \mathbf{x} - \boldsymbol{\zeta}(\mathbf{x}),
	\label{eq:lens_equation}
\end{equation}
where $\mathbf{y}$ represents the unlensed source position, 
$\mathbf{x}$ represents the corresponding lensed image position.
and $\boldsymbol{\zeta}(\mathbf{x})$ denotes the
deflection angle due to the presence of the lens.
The deflection angle $\boldsymbol{\zeta}\left(\mathbf{x}\right)$
depends on the lens potential $\psi(\mathbf{x})$ as: 
\begin{equation}
	\boldsymbol{\zeta}(\mathbf{x}) = \frac{D_{\rm ds}}{D_{\rm s}}
	\nabla\psi(\mathbf{x}); \quad 
	\psi(\mathbf{x}) = \frac{1}{\pi}\int d^2x' \: \kappa(\mathbf{x}') 
	\: \ln|\mathbf{x-\mathbf{x}'}|,
	\label{eq:deflection_angle}
\end{equation}
where 
\begin{equation}
	\kappa(\mathbf{x}) = \frac{\Sigma(\mathbf{x})}{\Sigma_{\rm cr}}, 
	\quad \Sigma_{\rm cr} = \frac{c^2}{4\pi {\rm G} D_{\rm d}}.
	\label{eq:convergence}
\end{equation}
The convergence $\kappa(\mathbf{x})$ represents the dimensionless 
surface mass density of the lens and $\Sigma_{\rm cr}$ denotes the
critical density for a source at infinity.
The $D_{\rm d}$, $D_{\rm s}$, and $D_{\rm ds}$ are the angular
diameter distances from observer to lens, observer to source and
from lens to source, respectively.

Various properties of the lens equation can be described by 
the corresponding Jacobian matrix:
\begin{equation}
	\begin{split}
	\mathbb{A}(\mathbf{x}) = \frac{\partial\mathbf{y}}{\partial\mathbf{x}} =
	\left(\begin{array}{cc} 1 & 0 \\
	0 & 1 \end{array}\right) - \frac{D_{\rm ds}}{D_{\rm s}}
	\left(\begin{array}{cc} \psi_{11} & \psi_{12} \\ 
	\psi_{21} & \psi_{22} \end{array}\right) \\
	=
	\left(\begin{array}{cc} 1 & 0 \\
	0 & 1 \end{array}\right) - \frac{D_{\rm ds}}{D_{\rm s}}
	\left(\begin{array}{cc} \kappa + \gamma_1 & \gamma_2 \\ 
	\gamma_2 & \kappa-\gamma_1 \end{array}\right)
	\end{split}
	\label{eq:jacobian}
\end{equation}
where $\psi_{ij}$ represents the second derivatives of the potential
and known as the \textit{deformation tensor}.
In the above equation, we have introduced the convergence ($\kappa$) 
and shear ($\gamma \equiv \sqrt{\gamma_1^2 + \gamma_2^2}$) which 
can be written in terms of the second derivatives of the lens 
potential as
\begin{equation}
	\kappa = \frac{1}{2}(\psi_{11} + \psi_{22}); \quad
	\gamma_1 = \frac{1}{2}(\psi_{11} - \psi_{22}); \quad
	\gamma_2 = \psi_{12}.
\end{equation}
The convergence ($\kappa$) introduces an isotropic distortion in 
the lens images whereas the shear stretches the lensed image in 
one particular direction and compresses in other.

The magnification factor of an lensed image is given as:
\begin{equation}
	\mu = \frac{1}{\det\mathbb{A}} = \frac{1}{(1-a\alpha)(1-a\beta)},
	\label{eq:mag_factor_pnt}
\end{equation}
where $a=D_{\rm ds}/D_{\rm s}$ is the distance ratio and $\alpha$
and $\beta$ are the eigenvalues of the deformation tensor.
The above equation for the magnification factor is valid for a 
point source. 
For an extended source, we have to take a weighted average over 
the source area:
\begin{equation}
	\mu(\mathbf{y}) = \frac{\int \mu_{\rm p}(\mathbf{y}')
	I(\mathbf{y}-\mathbf{y}')d^2\mathbf{y}'}{\int I(\mathbf{y}')
	d^2\mathbf{y}'}
	\label{eq:mag_factor_ext}
\end{equation}
where $\mu_{\rm p}$ represents the point source magnification 
(Equation \ref{eq:mag_factor_pnt}) and $I(\mathbf{y})$ represents
the surface brightness profile of the source.

The magnification for a point source (in principle) goes to infinity
at point in the image plane where $\alpha = 1/a$ or $\beta = 1/a$ or
both $\alpha = 1/a = \beta$.
These points with the infinite magnification are known as 
singularities of the lens mapping and form smooth closed curves 
in the image plane, known as \textit{critical curves}.
The corresponding points in the source plane (also closed curves
but not necessarily smooth) are known as \textit{caustics}.
The critical curves can be further divided into two categories: 
tangential and radial critical curves. 
For simple lens models, one can easily distinguish between two 
types as the images forming near tangential (radial) critical 
curves are tangentially (radially) elongated with respect to the 
lens center. 
The corresponding caustics are known as tangential and radial caustics. 

As discussed in \citetalias{2020MNRAS.492.3294M}, there are two types of 
singularities in gravitational lensing, namely, stable and unstable
(also knowm as point singularities).
Fold and cusp are known as stable as they are present for all 
possible source redshift.
On the other hand, for example, lips, beak-to-beak, swallowtail, 
and umbilics are unstable singularities as they are only present for
specific source redshifts and are very sensitive to the lens 
parameters.

The set of points in the image plane, which correspond to 
cusps in the source plane (for all possible source redshifts), 
form $A_3$-lines in the image plane.
These $A_3$-lines are the backbone of a singularity map (please
see \citetalias{2020MNRAS.492.3294M} for more details about the
singularity map) and all other point singularities lie on them.
In the image plane, $A_3$-lines locate the points where the gradient 
of the deformation tensor eigenvalue is orthogonal to the 
corresponding eigenvector: $n_\lambda . \nabla_x\lambda = 0$.
As deformation tensor has two eigenvalues ($\alpha$ and $\beta$), 
there are two set of $A_3$-lines corresponding to cusps 
on tangential and radial caustics in the source plane. 
Along with satisfying the $A_3$-line condition, point singularities
also satisfy additional criteria.
Swallowtail singularity indicate the points where eigenvector 
$n _\lambda$ of the deformation tensor is tangent to the 
corresponding $A_3$-line.
The umbilics denote degenerate points where different $A_3$-lines meet
with each other. At hyperbolic (elliptic) umbilic two (six)
$A_3$-lines meet with each other: one (three) corresponding to 
$\alpha$ and one (three) corresponding to $\beta$ eigenvalue.

As shown in \citetalias{2021MNRAS.503.2097M} and
\citetalias{2021arXiv210313617M}, a singularity map consisting of 
$A_3$-lines and point singularities is a very compact 
representation of the lens.
It marks all the regions in the lens plane with high magnification
and makes the comparison of different (real or simulated) lenses 
very simple.

\section{HFF Clusters}
\label{sec:hff_clusters}

\begin{figure*}
	\centering
	\includegraphics[height=8cm, width=17cm]
	{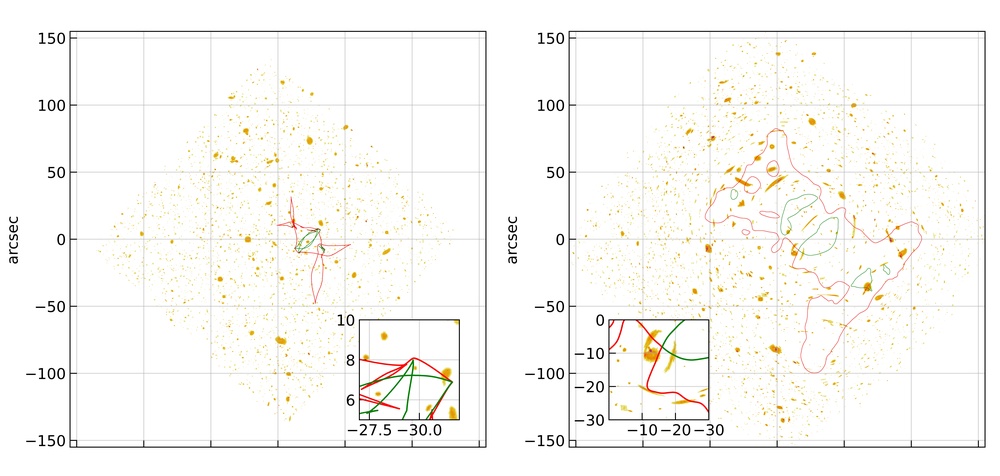}
	
	\vspace{-0.5cm}
	
	\includegraphics[height=8cm, width=17cm]
	{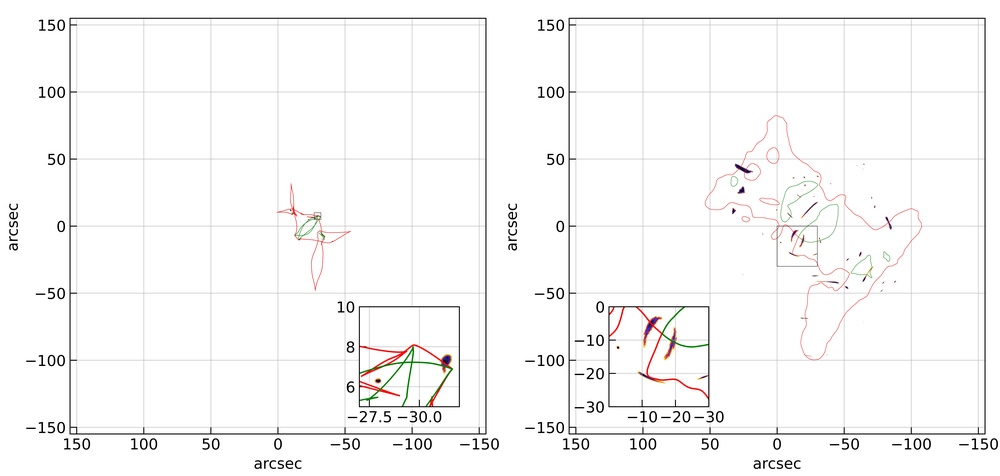}
	\caption{An example of lensed HUDF with purse-like image formation: 
	The left and right panel in top row represent all the sources from
	\citetalias{2020arXiv201209994N} in the unlensed and lensed (due to 
	A370 cluster) HUDF. The red and green lines represent the caustics 
	and critical lines in the left and right panel for a source redshift 
	($z_{\rm s}$) 3.7, respectively. The insets in left snd right panels 
	represent the 
	zoomed in view of unlensed and lensed source position in the source 
	and lens plane, respectively. In the bottom row, all the sources 
	below magnification ($\mu$) 30 are removed in order to clearly observe 
	the	characteristic image formation near purse image formation.} 
	\label{fig:a370_real_4}
\end{figure*}

The HFF clusters are six massive merging galaxy clusters targeted 
by the Hubble Space Telescope (HST) under the Hubble Frontier Fields 
(HFF) survey. 
The extensive study of these HFF clusters has allowed us to better
understand the high redshift distribution of galaxies,
\citep[e.g.,][]{2015MNRAS.450.3032M, 2018MNRAS.479.5184A}
the nature of dark matter \citep[e.g.,][]{2016MNRAS.458..660H, 
2016MNRAS.463.3876J, 2017ApJ...851...81A, 2017ApJ...836...61M},
the intracluster medium \citep[][]{2017ApJ...846..139M, 
2018MNRAS.474..917M} and others.
For each HFF cluster, various groups reconstructed the mass models
using parametric \citep[e.g.,][]{2009ApJ...703L.132Z, 
2010PASJ...62.1017O, 2014MNRAS.443.1549J, 2014ApJ...797...48J}, 
non-parametric \citep[e.g.,][]{2016MNRAS.459.1698M, 2018ApJ...868..129S,
2018MNRAS.480.3140W}, 
and hybrid techniques \citep[e.g.,][]{2014MNRAS.437.2642S}. 
Due to the different sets of assumptions considered by different 
teams, the final best-fit mass models for a given cluster lens 
can show considerable differences from each other
\citep[e.g.,][]{2017MNRAS.472.3177M, 2017MNRAS.465.1030P, 
2020MNRAS.494.4771R}. 
A difference in singularity maps corresponding to parametric and 
non-parametric mass models for the HFF clusters has been discussed 
in \citetalias{2021MNRAS.503.2097M}.
Comparing these different mass models for a particular cluster shows 
that the best-fit mass models corresponding to non-parametric 
reconstruction method \textsc{grale} lead to the simplest lens 
mass models and singularity map even if we include the statistical
uncertainties (please see \citetalias{2021arXiv210313617M}).
Hence, the best-fit \textsc{grale} mass models provide the lower
limit on the point singularity cross-section and three-image
arc (source lying near a cusp) cross-section.

In current work, we use the best-fit \textsc{grale} mass 
models (v4) for all of the HFF clusters as our lens mass models.
As mentioned above, this provides the most conservative estimates.
The resolution of these mass models has been fixed to $0.06''$.
As the best-fit \textsc{grale} mass models do not have significant
power in galaxy scale structures, a resolution of $0.06''$ is
adequate for our work.
The relevant lensing quantities (e.g., lens potential, deflection
angle, magnification) are calculated over a grid of 
$5270 \times 5270$ pixels $(316.2'' \times 316.2'')$.
Considering such a large angular region is necessary to cover 
all the sources in the HUDF region (discussed in the following 
section).
These best-fit \textsc{grale} mass models of the HFF clusters 
are obtained using only the strong lensing data. 
Hence, it is possible that the regions far from the 
cluster center are not very well constrained (see 
\citealt{2020MNRAS.494.3253L} for inclusion of weak lensing 
information in \textsc{grale}). 
However, it does not affect our work as we only focus on the 
strong lensing regions in all of the HFF clusters.

\section{HUDF}
\label{sec:hudf}

\begin{figure*}
	\centering
	\includegraphics[height=8cm, width=17cm]
	{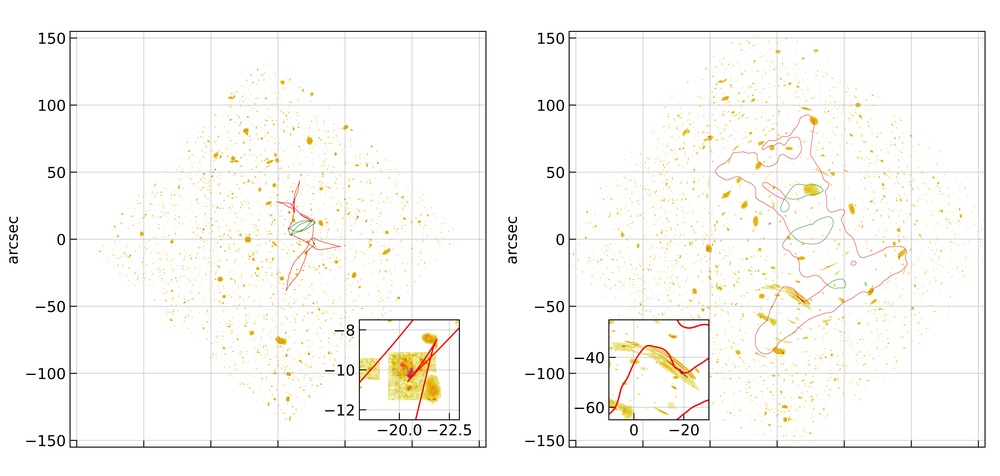}
	
	\vspace{-0.5cm}
	
	\includegraphics[height=8cm, width=17cm]
	{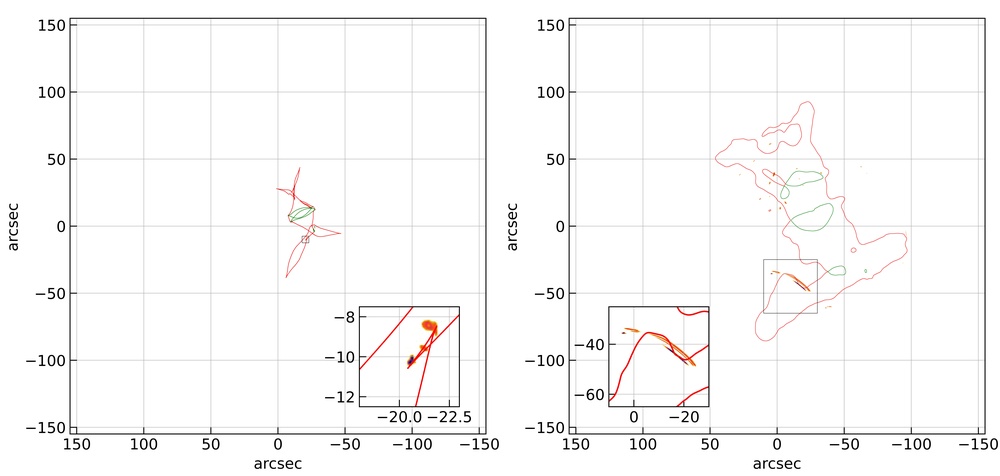}
	\caption{An example of lensed HUDF with swallowtail-like image formation: 
	The left and right panel in top row represent all the sources from
	\citetalias{2020arXiv201209994N} in the unlensed and lensed (due to 
	A370 cluster) HUDF. The red and green lines represent the caustics 
	and critical lines in the left and right panel for a source redshift 
	($z_{\rm s}$) 2.5, respectively. The insets in left snd right panels 
	represent the 
	zoomed in view of unlensed and lensed source position in the source 
	and lens plane, respectively. In the bottom row, all the sources 
	below magnification ($\mu$) 30 are removed in order to clearly observe 
	the	characteristic image formation near purse image formation.} 
	\label{fig:a370_real_1}
\end{figure*}

The Hubble Ultra Deep Field \citep[HUDF;][]{2006AJ....132.1729B} 
is the deepest image of the sky taken by the Hubble Space 
Telescope (HST) over a time equivalent to ${\sim}22$ days over 
the years reaching the magnitude limit ($m_{AB}$) ${\sim}30$.
The HUDF has a patch size of ${\sim}12 \: {\rm arcmin}^2$ 
containing ${\sim}10^4$ galaxy sources in the (photometric) 
redshift range from ${\sim}\left[0.02, 10\right]$ 
\citep{2015AJ....150...31R}.
So far, the Hubble Deep Fields have been used in many studies, 
like, galaxy luminosity function at high redshifts
\citep[e.g.,][]{2013ApJ...773...75O}, 
the evolution of galaxy properties with redshift
\citep[e.g.,][]{2006ApJ...647..787T, 2013ApJ...777..155O}, 
weak lensing studies 
\citep[e.g.,][]{2007ApJ...671.1182I, 2010ApJ...713..603B}
leading to better understanding of the Universe and its properties.

Apart from HUDF, there are other deep field observations, like,
the Great Observatories Origins Deep Survey 
\citep[GOODS;][]{2004ApJ...600L..93G},
the Galaxy Evolution from Morphologies and SEDs
\citep[GEMS;][]{2004ApJS..152..163R},
and the Cosmological Evolution Survey
\citep[COSMOS;][]{2007ApJS..172....1S}.
These surveys cover a significantly large area 
(GOODS: ${\sim}320{\rm arcmin}^2$, GEMS: ${\sim}800{\rm arcmin}^2$
COSMOS: ${\sim}2{\rm deg}^2$) of the sky compared to the HUDF.
Although such large area coverage helps us to counter the 
cosmic variance, but, limits the depth of the survey.
Keeping the unprecedented depth of the HUDF in mind, we
proceed with HUDF in this work and leave the other surveys
for future work.
In addition, the HUDF appears to be relatively under-dense as 
compared to wider deep fields like COSMOS at shallow magnitudes,
$m_{\rm AB}{\sim}[22,24]$ \citep{2007ApJS..172..219L}. 
Therefore, it is likely that we err towards an under-estimate of
cross-section.
This is in keeping with all our assumptions where we stay with a
lower bound on cross-section.

As the source galaxies in the HUDF lie at different redshifts, 
one need to extract them using software like \textsc{SourceExractor} 
\citep{1996A&AS..117..393B} from the patch in order to lens them.
In our work, we use the HUDF source galaxy cutouts (extracted using
\textsc{SourceExractor}) from 
\citet[][hereafter N20]{2020arXiv201209994N} as source templates.
Following \citetalias{2020arXiv201209994N}, we use only nearly half 
(No. of sources = 5271) of the brightest sources from the HUDF in our work.
These sources are extracted from the HUDF with a resolution of
$0.03''$ in four different filters B, V, i, z with the central
wavelength ${\sim}4320$\AA, ${\sim}5920$\AA, ${\sim}7690$\AA, 
${\sim}9030$\AA, respectively, and are provided in RGB image format
(please see \citetalias{2020arXiv201209994N} for more details).
It is important to note that by restricting ourselves to a 
subset of galaxies in the HUDF as potential sources, we are once 
again under-estimating the lensing cross-section.  
This is in the spirit of working with lower limits as in
\citetalias{2021MNRAS.503.2097M}.

\section{Results}
\label{sec:results}

\begin{figure*}
	\centering
	\includegraphics[height=6cm, width=8.5cm]
	{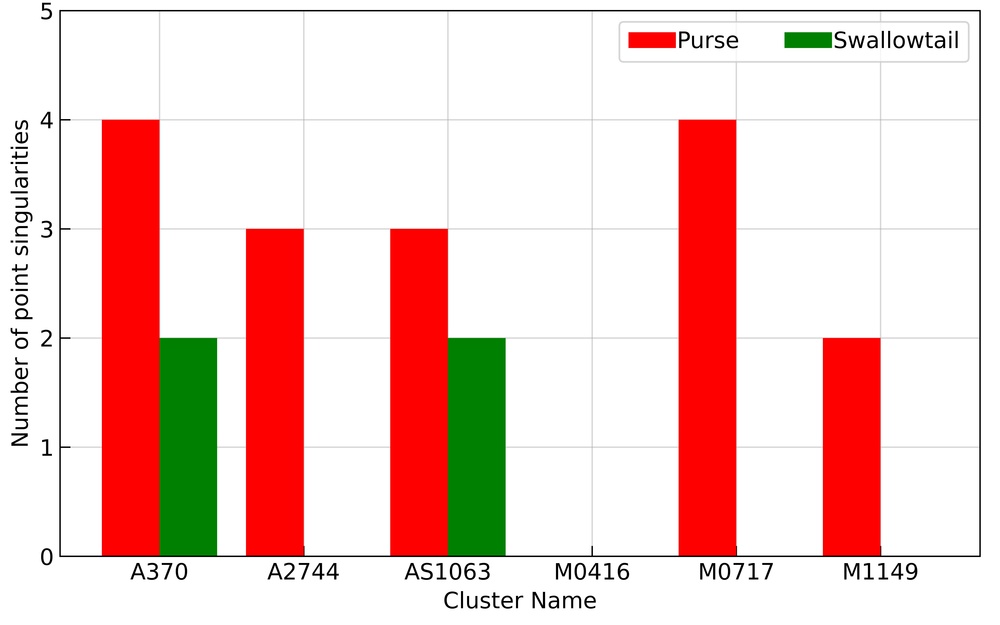}
	\includegraphics[height=6cm, width=8.5cm]
	{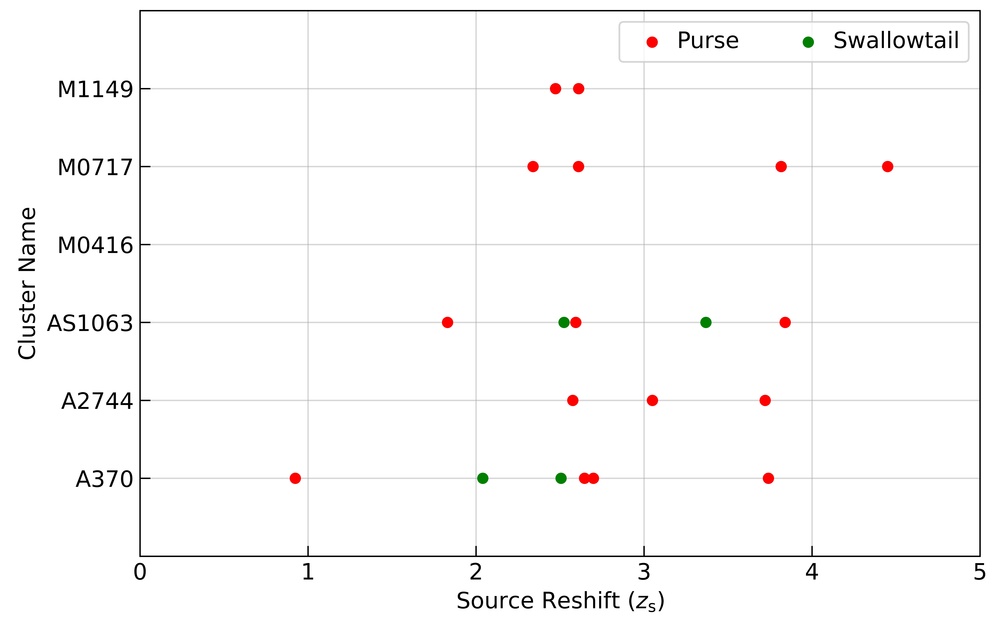}
	\caption{Distribution of identified exotic image formation in 
	HFF clusters: The left panel represents the number of exotic image 
	formations identified in the different HFF clusters. The red 
	and green bars represent the number such image formations of 
	purse and swallowtail singularities, respectively. The right 
	panel depicts the corresponding redshift distribution. Again 
	the red and green points represent the purse and swallowtail 
	singularities, respectively.} 
	\label{fig:hff_sing}
\end{figure*}

In this section, we present our results of lensing of the HUDF
by the HFF clusters.
In Subsection \ref{ssec:method}, we discuss our method in detail.
In Subsection \ref{ssec:exotic_purse} and \ref{ssec:exotic_swallow}, 
we discuss the number of exotic image formation in the lensed HUDF.
In Subsection \ref{ssec:time_delay}, we estimate the time delay
between multiple images in these exotic images.

\subsection{Methodology}
\label{ssec:method}

Once we get cutouts of the 5271 galaxy sources in the HUDF 
from \citetalias{2020arXiv201209994N}, we re-scale every source
to the resolution of lens plane (i.e., $0.06''$).
Doing such a rescaling smears the small-scale details in 
the source galaxies. 
However, as we mainly focus on exotic image formation in the
present work, it does not affect their identification (or the
corresponding cross-section).
After that, we lensed every source whose redshift is greater 
than the refshift of the lens.
Doing this give us a different number of lensed sources
for different HFF clusters.
A faster way to lens such a large number of lenses is to divide 
the sources into redshift bins and lens all the sources in one 
redshift bin simultaneously.
Such a method is more useful in generating a large number of
simulated sky patches \citep[e.g.,][]{2016ApJ...828...54L} which
is not the case here.
Hence, in our current work we lens every source individually.
As earlier work suggests \citep{2007ApJ...671.1182I, 2013MNRAS.435..822R}, 
the HUDF does not contain any strong 
lensing system, and only weak lensing signals are present. 
Hence, one can simply use one lens plane with multiple source 
plane configuration.

As the Einstein radius ($\theta_{\rm E}$) of the HFF clusters
at source redshift 9 is ${\sim}40''-60''$ 
\citep{2019MNRAS.486.5414V}, the area of the 
HUDF is considerably larger than the strong lensing region
of the HFF clusters.
Hence, we make 11 lensing realizations for each of the HFF clusters. 
In the zeroth realization, the center of the HUDF and the 
cluster lens center are aligned with each other and no rotation is
introduced. 
The HFF cluster center is the same as the center of v4 mass models
of the Williams group.
For rest of the ten realizations, we randomly vary the lens 
center within 1 ${\rm arcmin}^2$ square region centered at the 
HUDF center and rotate in by an arbitrary angle.
The size of this region (for cluster center variation) is chosen 
such that the strong lensing region always remains within the HUDF
for a source redshift of ten.

As the characteristic image formation takes place when the source 
is lying near the caustic, the corresponding magnification is higher. 
Hence, in such a large source population, we use different 
magnification cuts to remove undesirable sources. 
Removal of these sources allows us to easily identify the 
characteristic image formations near point singularities.
An example of such a magnification cut is shown in Figure
\ref{fig:a370_real_0}. 
The left panel represents the unlensed HUDF with all the sources.
The middle panel shows the corresponding lensed counterpart
with all the sources.
Here A370 has been used as the lens and the centers of both
A370 and HUDF are aligned with each other without any arbitrary 
rotation of the lens, i.e., the zeroth realization.
As expected the area covered by all the sources in the image 
plane has been increased after lensing and only the central
region shows the strong lensing properties.
The right panel represents the same lensed counterpart of left
panel but only with the sources that are magnified such that
$\mu{\geq}5$.
As mentioned above, doing so removes a significant fraction
of sources and allow us to easily see  only the highly magnified 
sources.
In actual observation, the availability of spectroscopic 
information can be used to distinguish different counterparts of 
the lensed source and identify the possible characteristic image 
formation. 
However, as we are only looking at the image formations in our 
current work, applying different magnification cuts is the only 
(practical) way to identify the characteristic image formations.

Once we identify and locate image formations, we 
find out the corresponding source redshift. 
After finding the source redshift, we replot the lens and source 
plane only for that particular source along with the critical 
curves and caustics. 
As we know the caustic evolution near point singularities, by doing 
this, we can confirm whether the image formation is due to 
a point singularity or not.
We would like to point out that we have used the second step to
eliminate some image formations that are chosen from visual
inspection.

\subsection{Exotic Image Formation: Purse}
\label{ssec:exotic_purse}

\begin{figure*}
	\centering
	\includegraphics[height=6cm, width=8.5cm]
	{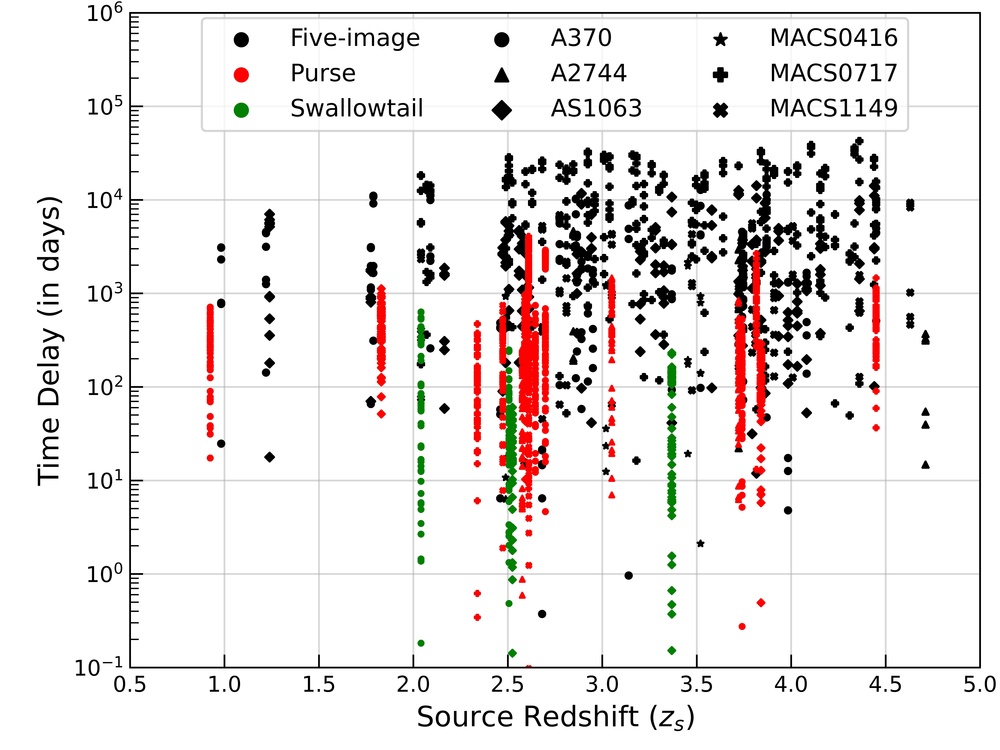}
	\includegraphics[height=6cm, width=8.5cm]
	{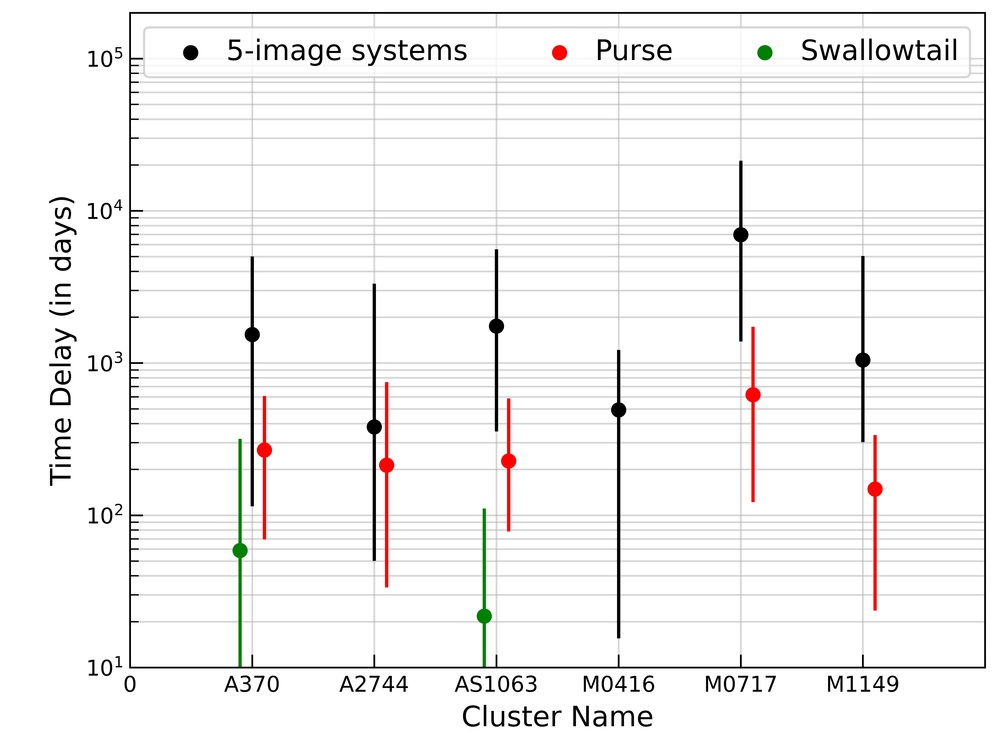}
	\caption{Time delay near point singularities: In the left 
	panel, the x-axis represents the source redshift and the
	y-axis represents the time delay in days. The black points
	represents the time-delay corresponding to typical 
	five-image geometry in all realizations for all of the HFF
	clusters. The red and green lines represent the time delay
	corresponding to the purse (hyperbolic umbilic) and 
	swallowtail. In the right panel, the black, red and green 
	points represent the median value of five-image systems, 
	purse and swallowtail for each HFF cluster.The scatter 
	around median value is represented by the $16{\rm th}$ and 
	$84{\rm th}$ percentile values.} 
	\label{fig:td_hff}
\end{figure*}

One example of the above-mentioned scheme is shown in Figure 
\ref{fig:a370_real_4} for an image formation near purse singularity. 
Here, the HUDF is lensed due to A370. 
In this particular case, the center of the HUDF and A370 are not 
aligned with each other and the lens is rotated by an arbitrary angle. 
In the top row of Figure \ref{fig:a370_real_4}, the left and right 
panels represent the unlensed and lensed HUDF images with all of the
sources (with $z_{\rm s} {>} z_{\rm L}$) from \citetalias{2020arXiv201209994N}. 
The red and green lines represent the caustics and critical curves
in the left and right panel, respectively for a source at redshift
($z_{\rm s}$) 3.7 which is the redshift of the source that is giving
rise to the characteristic image formation.
Again as expected, only the sources near the center of the cluster
lens are strongly lensed and others are only weakly lensed
and that the lensed HUDF patch covers more area than 
the unlensed HUDF patch.

In Figure \ref{fig:a370_real_4}, one can see that the exotic 
image formation near purse singularity has a small overlap 
with a different source. 
Significant overlap with other sources can lead to additional 
challenges in the identification of exotic image formation near 
point singularities. 
As mentioned above, one can deal with such difficulties if 
spectroscopic observations of the sources are available. 
In our current work, we apply magnification cuts on the 
sources in order to remove the unrequired sources and possible
overlap with exotic image formation.
The bottom row in Figure \ref{fig:a370_real_4}, represents the
top row with a threshold magnification ($\mu_{\rm th}$) of 30.
One can see that this removes a large fraction of the sources
along with the overlapping source and one can clearly see the
exotic image formation (ring like structure) near purse singularity.
The inset plots show the corresponding zoomed regions near the 
source and characteristic image formation in the source and lens 
planes.

The other important feature observed in this particular 
realization is the fact that only a part of the source (which is
inside the both radial and tangential caustics) is responsible 
for the exotic image formation which leads to the observations of 
merging images in these exotic image formations.
One would have observed four separate images if the complete 
source was within the caustics. 
The current image formation shows one possible variation of the
purse image formation.
So far (best to our knowledge), only one image formation near
purse is observed \citep{2008A&A...489...23L, 2009MNRAS.399....2O}
in cluster lens Abell 1703.

\subsection{Exotic Image Formation: Swallowtail}
\label{ssec:exotic_swallow}

Another example of exotic image formation near swallowtail 
singularity is shown in Figure \ref{fig:a370_real_1}. 
Here again, the cluster lens is A370. 
However, the lensing realization is different from Figure 
\ref{fig:a370_real_4} meaning that the lens alignment and 
rotation angle is different.
Again, the top row in Figure \ref{fig:a370_real_1} shows 
the unlensed and lensed HUDF in the left and right panel, 
respectively, with all the sources having source redshift
$z_{\rm s} {>} z_{\rm L}$ . 
The red and green lines represent the caustics and critical 
curves in the left and right panels, respectively, for a source 
redshift ($z_{\rm s}$) of 2.5. 
The bottom row is the same as the top row but with a threshold 
magnification ($\mu_{\rm th}$) of 30.
As the sources are extended galaxy sources, one can see that
only a handful of source are left with a magnification $> 30$.
The inset plots show the corresponding zoomed regions near the 
source and characteristic image formation in the source and lens 
planes.

As we know from \citetalias{2020MNRAS.492.3294M}, the characteristic
image formation for a swallowtail singularity is an arc made of
four images which we also observe in the \ref{fig:a370_real_1}.
However, in Figure \ref{fig:a370_real_1}, instead of having an 
overlap of multiple sources (like Figure \ref{fig:a370_real_4}), 
we observe three different sources lying near the swallowtail 
singularity and giving rise to corresponding image formation. 
The galaxy source in the middle of the inset plot is at redshift 
2.5, and the bottom (top) source lies at a redshift 2.62 (2.86). 
Such image formations where several sources are lying near 
the point singularity can be observed for both swallowtail and 
purse. 
However, as the probability of observing two sources near 
each other in the sky as well as in redshift is small, one does 
not expect to such juxtaposition very often.
Observations have led us to the detection of a handful of cases 
with image formation near swallowtail singularity in both 
galaxy \citep[e.g.,][]{2010A&A...524A..94S} and cluster scale lenses
\citep[e.g.,][]{1998MNRAS.294..734A}.

In Figures \ref{fig:a370_real_4} and \ref{fig:a370_real_1}, 
we only showed one example of exotic image formation near both 
purse and swallowtail singularities with A370 as the cluster lens.
However, there are other realizations where we have identified
the image formation near these point singularities.
All these image formations (including from Figure \ref{fig:a370_real_4}
and \ref{fig:a370_real_1}) are available as supplementary
material. 
In Figure \ref{fig:hff_sing}, we show the number and redshift 
distribution of all exotic image formations identified in the 
HFF cluster lenses.
The left panel represents the number distribution of the exotic 
image formations in the HFF clusters, whereas the right panel 
represents the corresponding redshift distribution.
In the eleven realizations corresponding to the A370, we identified 
a total of four image formations near the purse and three near the 
swallowtail singularity. 
On the other hand, we did not observe any exotic image formation 
in M0416 realizations.
This again backs up the earlier result of \citetalias{2021arXiv210313617M} 
that A370 (M0416) is the most (least) efficient in producing 
image formation near point singularities in all of the HFF clusters
considering the best-fit \textsc{grale} lens mass models.

From the redshift distribution of exotic image formations in the 
right panel, one can see that most of the images are identified in 
the redshift range [1, 4]. 
Such a result is expected as at the lower redshifts, the strong 
lensing region in the cluster is small, and at higher redshifts, 
the source density decreases.
However, we again remind the reader that here we are only 
considering half of the sources from HUDF. 
Increasing the number of sources increases the probability
of observing such image formation at both low and high redshift.

\subsection{Time Delay Analysis}
\label{ssec:time_delay}

Various point singularities mark the points in the source plane 
where cusp are created or destroyed or exchanged between radial 
and tangential caustic. 
Hence, the characteristic image formations are only observed if 
the source lies near the caustics in the source plane.
This implies that the time delay between various images, which 
are part of the characteristic image formation, should be smaller 
than a typical strong lensing scenario where the source lies
sufficiently far away from point caustics.

We calculate the time delay between different image pairs for 
each identified exotic image formation in all realization.
Figure \ref{fig:td_hff}, represents the time delay values in 
five-image geometry and in purse and swallowtail image 
configurations.
The left panel is a combined plot of the time delay for 
five-image (black points), purse (red points) and swallowtail 
(green points) geometry found in all realizations for all of the 
HFF clusters.
Here we removed the global minima image from all the systems.
Doing so leads to a total of six pairs of time-delays for one
five-image system.
As the number of image formations near purse and swallowtail is
very small compared to the number of typical five-image cases,
we draw a $5{\rm Kpc}$ circle around the identified purse and
swallowtail image formation and choose ten different random 
source position and calculate the corresponding time delays.
Doing so allow us to estimate the possible variation in the
time delay associated with one particular system.

One can see that the a typical five image geometry gives time
delay values in range ${\sim}[1, 50000]$ days.
On the other hand, image formations near purse and swallowtail
(except for a few geometries) always remain less than 1000 days
which is smaller than the typical five-image geometry
time delay.
This is evident from the right panel of Figure 
\ref{fig:td_hff}, where we show the median time delay values
for each HFF cluster separately. 
Again, one can see that the median value for five-image geometry
for each HFF is nearly an order of magnitude larger than the
purse and swallowtail geometry.
The error bars around these median points cover the range
between $16{\rm th}$ and $84{\rm th}$ percentile.

From left and right panels, one can also notice that, in
general, the swallowtail geometry leads to smaller time delays
compared to the purse geometry.
The time-delays corresponding to the pyramid singularity would
have been even smaller than the swallowtail but the \textsc{grale} 
best-fit mass models for HFF cluster only give one pyramid 
point (in MACS1149) leading to a negligible cross-section.
Hence, we do not study the image formation or time-delay
analysis for the pyramid singularities in our current work.
However, the same is not true for the parametric mass models.
As we have see in \citetalias{2021MNRAS.503.2097M}, the number
of pyramid singularities are significantly larger compared to
the non-parametric mass models encouraging us to include pyramid
in our future analysis.

\section{Conclusions}
\label{sec:conclusions}

In our current work, we have investigated lensed HUDF templates 
for exotic image formations near point singularities.
The best-fit \textsc{grale} mass models of the HFF cluster are 
chosen as the lens mass models as they provide the lowest 
cross-section for the point singularities 
(see \citetalias{2021MNRAS.503.2097M} and \citetalias{2021arXiv210313617M}).
We constructed eleven realizations of the lensed HUDF for 
every HFF cluster with different random orientations. 
The alignment of the HFF clusters and the HUDF is chosen such that
the critical lines for a source at redshift ten are always confined
within the HUDF region.
As the HUDF contains galaxy sources at various redshifts, the 
corresponding source cutouts (only for 5271  sources from
\citetalias{2020arXiv201209994N}) are used for lensing in current work. 
At present, we do not have an automated algorithm to locate 
these image formations. 
Hence, we visually inspected each of the realizations to 
identify the exotic image formation. 
Out of 66 realizations, in 4 (16) realizations, we identified 
image formations near the swallowtail (purse) singularity. 
In this work, we do not look for the image formation near a 
pyramid singularity as the best-fit \textsc{grale} mass models 
for the HFF clusters only give one pyramid singularity in MACS1149.

It is noteworthy that the number of realizations of purse singularity
is much higher than swallow tail.
This suggests that purse may be more common than swallow tail in deep
observations.
This possibility needs to be explored with a larger set of cluster
lenses.

Apart from image formation near point singularities, we also 
study the time delay between multiple images in these exotic
image formations. 
We find that typical time delay in swallowtail and purse
image formations are an order of magnitude smaller than a
typical five image geometry in the HFF clusters (after 
removing the global minima).
Such a difference can be very helpful in time-delay cosmography
to constrain the Hubble constant.

The key takeaway points of our current analysis is that the number
of image formation near point singularities are not negligible
whether we estimate the number using source galaxy population
or by using the simulated patches of the sky.
The corresponding time-delay values are significantly smaller 
than typical image formation in the cluster lenses.

Now that we have shown that the number of such systems is not 
small, the following questions need to be addressed: 
(i) What is the effect of the line-of-sight (off-plane)
structures on the number of point singularities for a given lens?
(ii) How to identify these image formations (including pyramids) 
in the all-sky surveys?
In order to account for the line-of-sight structure (if 
far away from the main lens), one needs to use double-plane 
(or multi-plane) lensing.
As discussed in the previous work \citep{1993A&A...272L..17L, 
1993A&A...278L..13K}, new type of singularities may also arise 
in the double plane lensing which can further complicate the analysis. 
Although the current analysis allows us to see variations in 
the image formations near point singularities, it is not complete.
The best-fit \textsc{grale} mass models underestimate the small
scale structures. 
Hence, the above identified image formations only cover a small 
fraction of possible variations in exotic image formation.
One can use both parametric and non-parametric mass models to
construct a catalog of possible image formations near point
singularity. 
Such a catalog can be helpful in automated searches in all-sky
surveys \citep[e.g.,][]{2018MNRAS.473.3895L, 2019ApJ...877...58A,
2019MNRAS.487.5263D}.
These problems are subjects of our future work, and the 
results will be presented in forthcoming publications.

\section{Acknowledgements}

AKM would like to thank Council of Scientific $\&$ Industrial 
Research (CSIR) for financial support through research 
fellowship  No. 524007.
Authors would like to thank Liliya Williams for providing lens 
mass models for the HFF clusters.
JSB would like to thank Yannick Mellier for suggesting the use 
of deep fields as sample source population.
This research has made use of NASA's Astrophysics Data System 
Bibliographic Service.
We acknowledge the HPC@IISERM, used for some of the computations 
presented here.


\section{Data Availability}

The high resolution HFF cluster mass models corresponding to 
the Williams group are available from the modelers upon 
request.


\bibliographystyle{mnras}
\bibliography{reference}



\bsp	
\label{lastpage}
\end{document}